\documentclass{emulateapj} 

\usepackage{graphics}

\shorttitle{Chandra LETGS observation of V4743}
\shortauthors{Ness, Starrfield, Burwitz, et al.}

\begin{document}
\title{A Chandra LETGS observation of V4743 Sagittarius: A Super Soft X-ray Source and a Violently Variable Light Curve}

\author{J.-U. Ness\altaffilmark{1}, S. Starrfield\altaffilmark{2},
V. Burwitz\altaffilmark{3}, P. Hauschildt\altaffilmark{1},
R. Wichmann\altaffilmark{1}, J.J. Drake\altaffilmark{4}, R.M.
Wagner\altaffilmark{5}, H.E. Bond\altaffilmark{6}, J.
Krautter\altaffilmark{7}, M. Orio\altaffilmark{8,9}, M.
Hernanz\altaffilmark{10}, R.D. Gehrz\altaffilmark{11}, C.E.
Woodward\altaffilmark{11}, Y. Butt\altaffilmark{4}, K.
Mukai\altaffilmark{12}, and S. Balman\altaffilmark{13}}

\altaffiltext{1}{Hamburger Sternwarte, Gojenbergsweg 112, 21029
Hamburg, Germany;\\ (jness, phauschildt, rwichmann)
@hs.uni-hamburg.de} \altaffiltext{2}{Department of Physics and
Astronomy, Arizona State University, Tempe, AZ 85287-1504:\\
starrfield@asu.edu} \altaffiltext{3}{Max-Planck-Institut f\"ur
extraterrestrische Physik, P.O. Box 1312, D-85741 Garching,
Germany:\\ burwitz@mpe.mpg.de}
\altaffiltext{4}{Harvard-Smithsonian Center for Astrophysics, 60
Garden Street, Cambridge, MA 02138:\\ jdrake@cfa.harvard.edu}
\altaffiltext{5}{Large Binocular Telescope Observatory, 933 North
Cherry Avenue,Tucson, Arizona 85721: \\ rmw@as.arizona.edu}
\altaffiltext{6}{Space Telescope Science Institute, 3700 San
Martin Dr., Baltimore, MD 21218: bond@stsci.edu}
\altaffiltext{7}{Landessternwarte K\"onigstuhl, D-69117
Heidelberg, Germany: jkrautter@lsw.uni-heidelberg.de}
\altaffiltext{8} {INAF-Turin  Observatory, Strada Osservatorio 20,
I-10025 Pino Torinese (TO), Italy: orio@to.astro.it}
\altaffiltext{9} {Department of Astronomy, University of
Wisconsin, 475 N. Charter St., Madison WI 53706} \altaffiltext{10}
{Instituto de Ciencias del Espacio (CSIC), \& Institut d'Estudis
Espacials de Catalunya, Edifici Nexus-201, C/ Gran Capit\`{a} 2-4,
E-08034 Barcelona, Spain: hernanz@ieec.fcr.es}
\altaffiltext{11}{Department of Astronomy, University of
Minnesota, 116 Church Street SE, Minneapolis, MN 55455:
\\ (gehrz, chelsea)@astro.umn.edu} \altaffiltext{12}{NASA Goddard
Space Flight Center, Code 662, Laboratory for High Energy
Astrophysics, Greenbelt, MD 20771: mukai@milkyway.gsfc.nasa.gov}
\altaffiltext{13} {Physics Department, Middle Eastern Technical
University, Inonu Bulvari 06531, Ankara, Turkey:
solen@astroa.physics.metu,edu.tr}

\begin{abstract}

V4743 Sgr (Nova Sgr 2002 No. 3) was discovered on 20 September
2002. We obtained a 5 ks ACIS-S spectrum in November 2002 and
found that the nova was faint in X-rays. We then obtained a 25\,ks
CHANDRA LETGS observation on 19 March 2003. By this time, it had
evolved into the Super Soft X-ray phase exhibiting a continuous
spectrum with deep absorption features. The light curve from the
observation showed large amplitude oscillations with a period of
1325\,s (22\,min) followed by a decline in total count rate after
$\sim$\,13\,ks of observations.  The count rate dropped from $\sim
40$\,cts s$^{-1}$ to practically zero within $\sim6$\,ks and
stayed low for the rest of the observation ($\sim6$\,ks.  The
spectral hardness ratio changed from maxima to minima in
correlation with the oscillations, and then became significantly
softer during the decay. Strong H-like and He-like lines of
oxygen, nitrogen, and carbon were found in absorption during the
bright phase, indicating temperatures between 1--2\,MK, but they
were shifted in wavelength corresponding to a Doppler velocity of
-2400\,km s$^{-1}$. The spectrum obtained after the decline in
count rate showed emission lines of C\,{\sc vi}, N\,{\sc vi}, and
N\,{\sc vii} suggesting that we were seeing expanding gas ejected
during the outburst, probably originating from CNO-cycled
material. An XMM-Newton ToO observation, obtained on 4 April 2003
and a later LETGS observation from 18 July 2003 also showed
oscillations, but with smaller amplitudes.
\end{abstract}


\section{Introduction}

Classical Nova explosions (CNe) occur in close binary systems
consisting of a main sequence star and a white dwarf (WD). Mass
transfer from the main sequence star onto the WD leads to
explosive hydrogen burning (a thermonuclear runaway: TNR) in the
hydrogen-rich matter accreted onto the surface of the WD if
ignition conditions are reached \citep{sf89, G98}.  CNe likely
recur on time scales $\sim10^5$\,yr depending upon the efficiency
of mass transfer in individual systems. CNe are classified as fast
or slow depending on the time to decline by three (visual)
magnitudes from maximum, $t_3$. A very fast nova will have
$t_3<20$\,days while a slow nova will have $t_3>100$\,days.
Multiwavelength observations of CNe in outburst provide information
on the energetics, the amount of mass ejected, and the chemical
composition of the ejecta. X-ray observations provide information
on the evolution of the underlying WD and the progress of the TNR.

Previous X-ray studies of novae {\it in outburst} were carried out
by EXOSAT, ROSAT, and most recently ASCA, BeppoSAX, and CHANDRA.
ROSAT detected V838 Her, V351 Pup, V1974 Cyg, GQ Mus, and LMC 1995
\citep{kr96, kr02, orio01}. The combined (BeppoSAX, ASCA, CHANDRA)
X-ray observations of V382 Vel show that it was initially a hard
source, evolved to a Super Soft X-ray Source (SSS) and then
declined extremely rapidly (factor of 200 in 6 weeks) to an
emission line source \citep{bw03}. In contrast, V1494 Aql was
observed to be an emission line source for about six months and
then evolved to a SSS. CHANDRA grating observations of this nova
in the SSS phase showed a spectrum qualitatively resembling that
of the SSS CAL 83 \citep{p01}. The light curve from this
observation exhibited a $\sim$2.5\,ks periodicity (which was
interpreted as pulsation) plus a factor of 6 burst \citep{drake}.

Nova V4743 Sgr was discovered by \cite{kats} at about m$_{\rm V} =
5$\,mag on 20 September 2002. \cite{west} provided an accurate
position of RA(J2000)$=19^{\rm h}01^{\rm m}09^{\rm s}.38$,
Dec(J2000)$=-22^\circ00\arcmin 05\arcsec .9$. The distance was
given by \cite{lyke} as $\sim$\,6.3\,kpc based on infrared
observations. The H$\alpha$ emission line was measured with a
full-width half-maximum (FWHM) of 2400\,km\,s$^{-1}$ \citep{kato}
implying ejection velocities exceeding 1200\,km\,s$^{-1}$. V4743
was classified as a very fast nova ($t_3<15$\,days) from the
visual light curve. In this Letter, we report on our observation
of this nova carried out with the CHANDRA LETGS.

\section{The Light Curve}

\begin{figure}[!ht]
\epsscale{1.2} \plotone{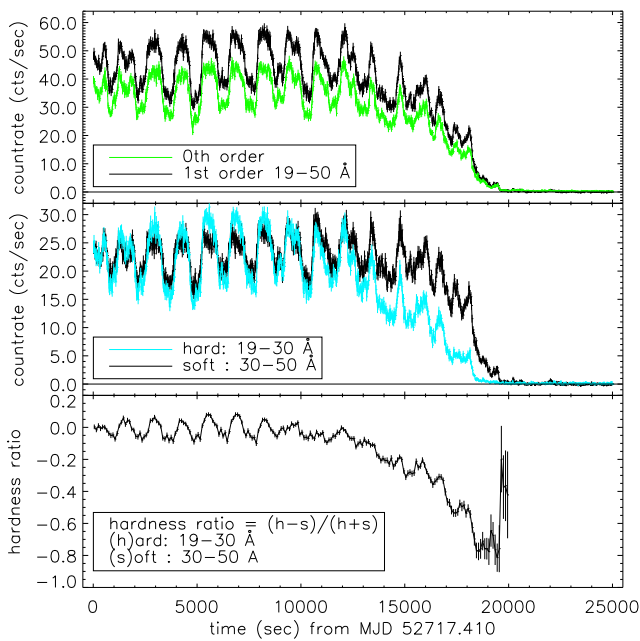} \caption{\label{lc25}Light curve of
the 25\,ks exposure (25\,sec bins) of V4743 Sgr extracted in the
designated wavelength intervals. {\bf Top:} Complete wavelength
ranges in 0$^{\rm th}$ and 1$^{\rm st}$ order. {\bf Middle:} Light
curve broken into ``hard'' and ``soft'' components of the
spectrum. {\bf Bottom:} Time evolution of the hardness ratio.}
\end{figure}

Our LETGS exposure was obtained on 19 March 2003 between
9$^h$30$^m$12$^s$ and 17$^h$01$^m$10$^s$ UT with a total exposure
time of 24.7\,ks (OBSID 3775). We extracted the dispersed spectrum
and the lightcurves (including removing the background photons)
using standard threads from CIAO (version 2.3:
http://asc.harvard.edu).

In Figure~\ref{lc25} (the top panel), we present the light curve extracted
from both the 0$^{\rm th}$ and 1$^{\rm st}$ orders. The nova is
extremely bright and we identify two important features of this
light curve. In the first phase of our observation, the count rate
shows strong variability with a range from $\sim$30-60\,cts
s$^{-1}$. About $\sim$13\,ks into the observation, a slow decline
starts and the light curve drops to nearly zero within
$\sim6$\,ks. In the middle panel of Figure~\ref{lc25}, we plot the light
curve extracted from ``soft'' (30--50\,\AA) and ``hard''
(19--30\,\AA) energy bands binned in 25s intervals. It can easily
be seen that the hard component declines first while the soft
component declines later but more rapidly. We also show this in
the bottom panel where we plot the time evolution of the spectral
hardness ratio defined by (h-s)/(h+s) with ``h'' the flux in
19\,\AA\,$<\lambda<30$\,\AA\ and ``s'' the flux in
30\,\AA\,$<\lambda<50$\,\AA.  The oscillations during the initial
phase can clearly be recognized as can the softening during the
decay phase.  We checked the observation and no known instrument
or satellite effect can provide a plausible explanation for the
decline.  We obtained an XMM ToO observation of this nova on 4
April, 2003 and a second LETGS spectrum on 18 July 2003 (to be
discussed elsewhere). The nova was bright and still oscillating in
both observations.

\begin{figure}[!ht]
\epsscale{1.2} \plotone{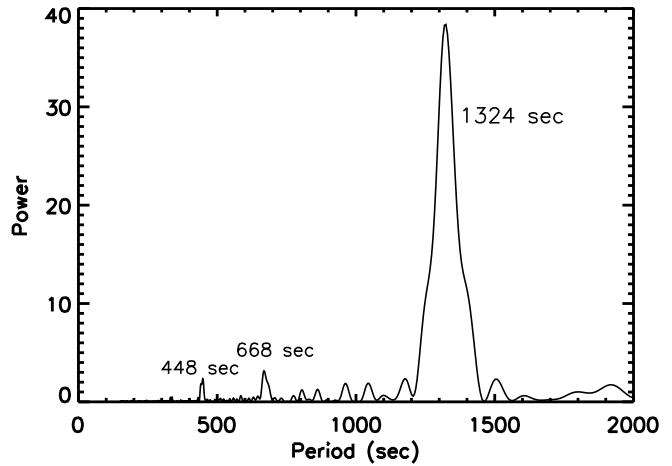} \caption{\label{per}Periodogram for
$0^{\rm th}$ order light curve with 22\,min period and two
harmonics marked. \vspace{-.3cm}}
\end{figure}

From the light curve we obtained a periodogram, plotted in
Figure~\ref{per}. We identify a strong period at 1324\,sec (22\,min). Two
harmonic overtone periods at 668\,sec and 448\,sec also appear in
the periodogram which is to be expected because of the complex
nature of the variations.

\section{The Spectrum}

\begin{figure}[!ht]
\epsscale{1.2} \plotone{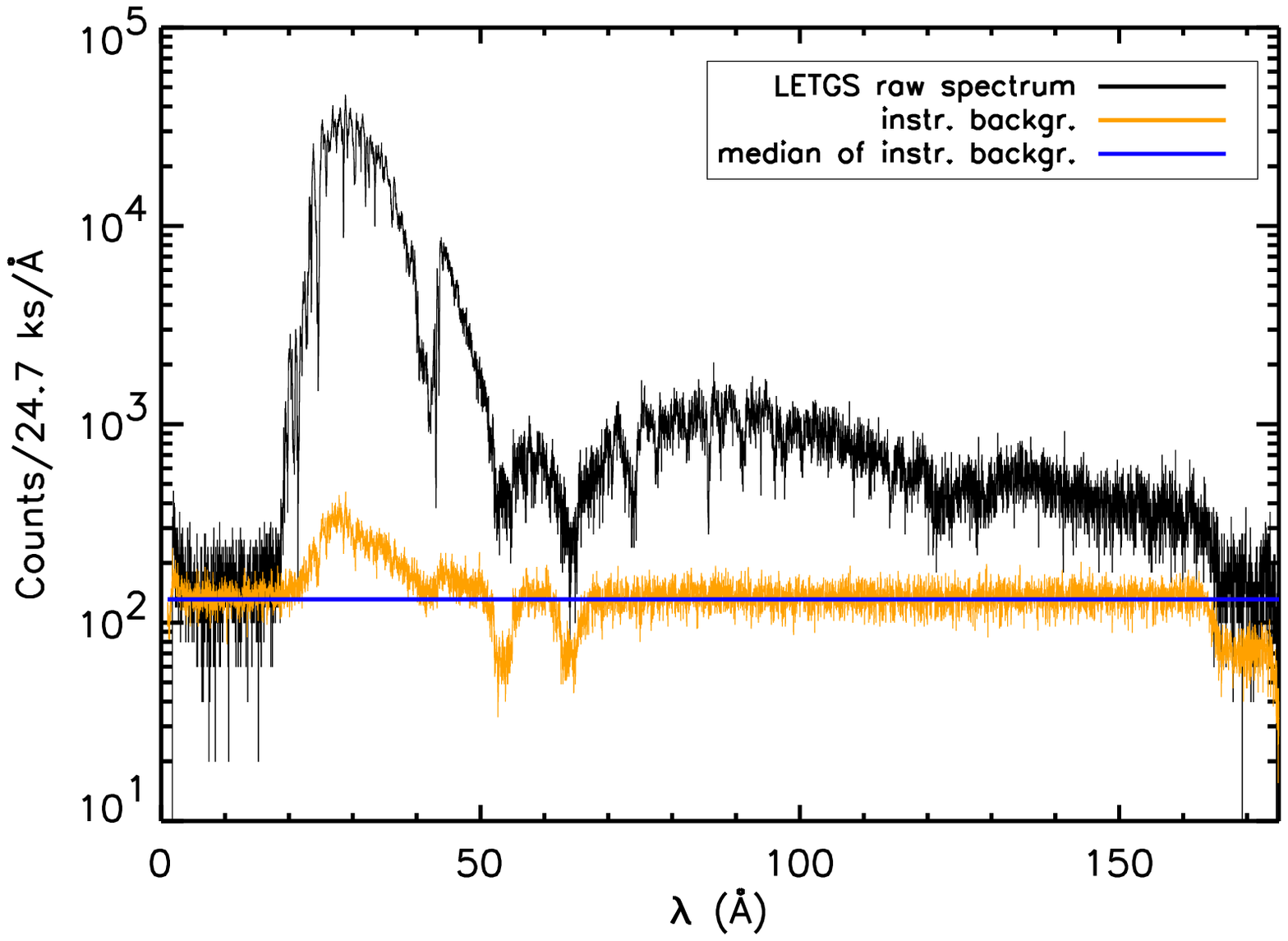} \plotone{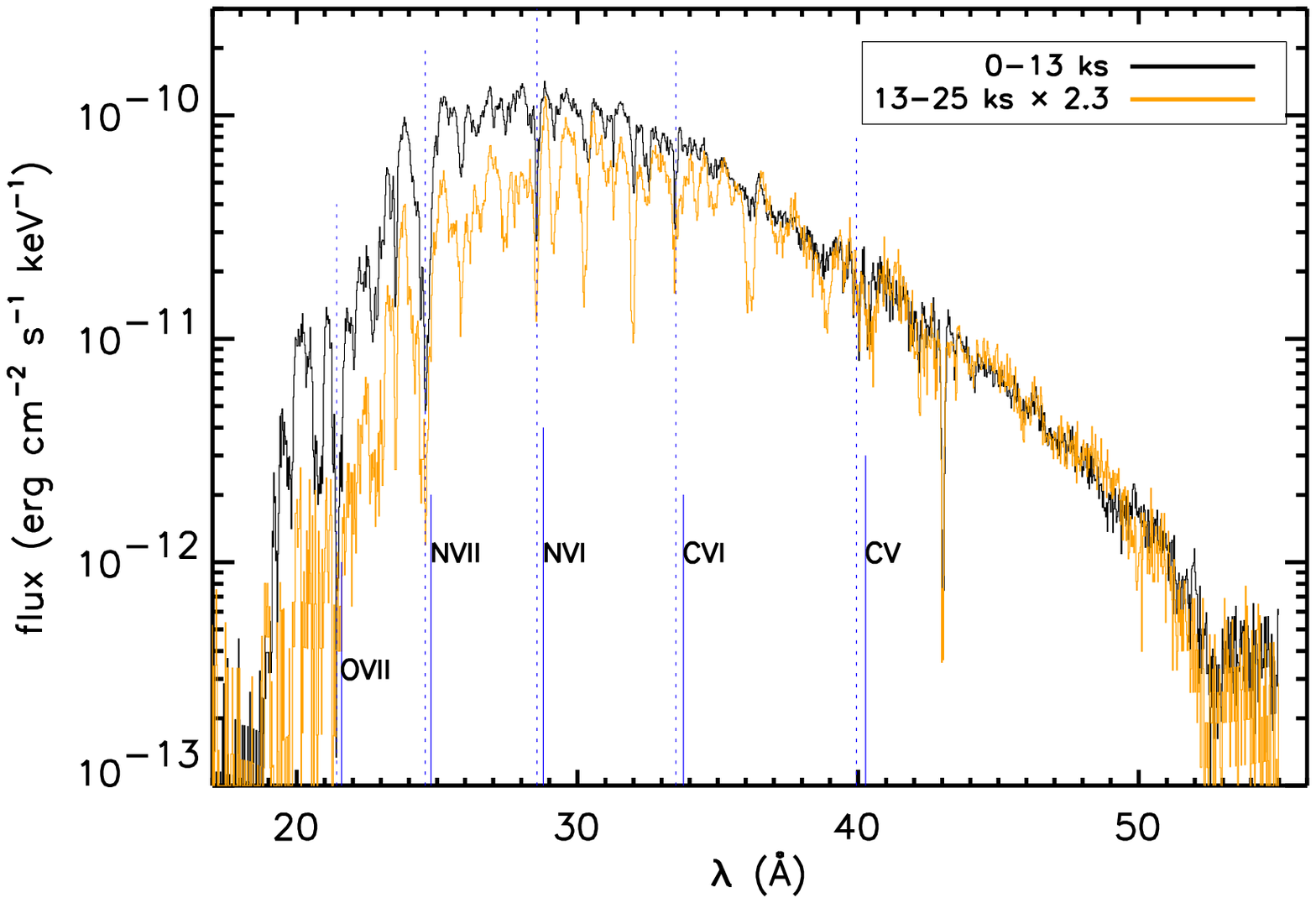}
\caption{\label{spec}{\bf Top:} Count spectrum (black) and
background spectrum (grey) with median background as black
horizontal line. {\bf Bottom:} Energy spectra (median-background
subtracted) for the first 13\,ks (black) and for the remaining
11.5\,ks multiplied with a factor 2.3 (grey). Identifications of
strong lines are given with vertical lines (solid: rest
wavelength; dotted: shifted by -2400\,km/s).}
\end{figure}

Figure~\ref{spec} shows the spectrum of V4743 displayed in two panels. The
top panel shows the count spectrum obtained from the entire 25\,ks
of our observation (both dispersion directions co-added). The edge
around 40\,\AA\ is an instrumental effect from absorption by
carbon in the detector coating (``carbon edge'').  A second broad
peak is visible at around 90\,\AA, but because the HRC detector
cannot disentangle higher order photons by their energy, we
modeled the higher orders (from 3--5) and found that this second
peak is caused by the overlapping higher orders of the 30\,\AA\
peak.  Since the background is otherwise flat, the median value
for all bins is a fair treatment of the instrumental background
(horizontal black line).

We obtained fluxes from the count spectrum, shown in the top
panel, by subtracting the median background and converting the
count numbers to photon energies using effective areas from
\cite{pease}. The fluxed spectrum is given in the bottom panel of
Figure~\ref{spec}. Because of the obvious structure in the light curve, we
display two spectra in this panel. The black line shows the
spectrum of the first 13\,ks of our observation, before the
beginning of the decline. The grey line over plots the spectrum of
the decline phase (the remaining 11.5\,ks multiplied by a factor
of 2.3 for better comparison).  The spectrum is clearly softer
during the decline phase than in the initial 13\,ks.

The spectrum is dominated by continuum emission from what appears
to be a stellar atmosphere with a peak $\sim$30\,\AA\ (0.4\,keV).
The continuum emission shows strong apparent ``absorption''
features. We tried to identify these features with strong lines
from the most abundant elements seen in nova ejecta such as
carbon, nitrogen, and oxygen \citep{bw03}. However, we could not
clearly identify any emission or absorption features associated
with the {\it rest} wavelengths for these lines (marked with solid
vertical lines in the bottom panel of Figure~\ref{spec}). We then applied
line shifts, using as a guide the velocity reported by
\cite{kato}, and found that we could identify the H-like and
He-like lines of nitrogen (N\,{\sc vii} at 24.78\,\AA\ and N\,{\sc
vi} at 28.78\,\AA) and carbon (C\,{\sc vi} at 33.8\,\AA\ and
C\,{\sc v} at 40.27\,\AA) as well as O\,{\sc vii} at 21.6\,\AA\
using a velocity of -2400 km s$^{-1}$. The blue shifted lines are
marked by dotted vertical lines in Figure~\ref{spec} The presence of these
ions implies a temperature of 1-2\,MK. The continuum temperature
will have to be obtained from stellar atmosphere modeling which is
underway. Note that our velocity is double the expansion velocity
found by \cite{kato} for the material ejected early in the
outburst.  We also extracted spectra during just the maximum and
minimum phases of the oscillations and found the continuum to be
slightly softer at minimum but the equivalent widths of the
absorption features are hardly changed.

\begin{figure}[!ht]
\epsscale{1.2} \plotone{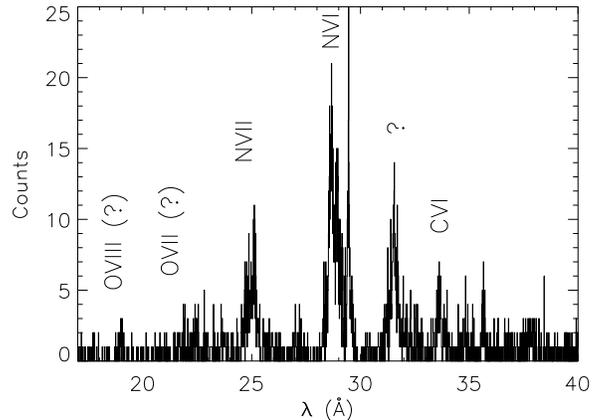} \caption{\label{l5}The spectrum
obtained during the last 5.4\,ks showing strong emission lines
from nitrogen and carbon.}
\end{figure}

In Figure~\ref{l5} we show the spectrum extracted during the last 5.4\,ks
of our observation, i.e., after the decay. While emission lines
from N\,{\sc vii}, N\,{\sc vi}, and C\,{\sc vi} can be identified,
there are also weak features that may indicate the presence of
O\,{\sc viii} (18.97\,\AA) and O\,{\sc vii} (21.6\,\AA). The
strong lines are broadened by $\sim 800-1200$\,km\,s$^{-1}$ and
probably arise in the ejecta. From the N\,{\sc vii}/N\,{\sc vi}
line flux ratio we infer a temperature of 1.6\,$\pm$\,0.1\,MK
using the atomic database APEC. The ratio of N\,{\sc vii}/C\,{\sc
vi} indicates an enhanced nitrogen abundance suggesting that we
are seeing CNO-cycled material \citep{cno}.

\section{Discussion}

This Letter reports on a CHANDRA LETGS+HRC observation of V4743
Sgr in outburst.  We find that the light curve exhibits large
amplitude oscillations.  However, the nova is only bright in
X-rays for the first 15 ks.  It then declines to nearly zero over
about 90 min and stays faint for the rest of the observation. The
behavior of the light curve is puzzling and, at this time, we can
only present two questions posed by this observation. The first is
are we viewing an eclipse or an actual decline of the X-ray
emission? The second is what is the cause of the oscillation:
pulsation or rotation?  The fact that XMM observed it about 3
weeks later and found it to be bright and oscillating suggests
that it was an eclipse.  However, the long duration of the decline
in count rate is difficult to interpret in the eclipse scenario.
If this system has a typical Cataclysmic Variable binary period of
a few hours, then the duration of the decline ($\sim$6\,ks) is too
long for an ingress into total eclipse of just the WD. There are
Cataclysmic Variables with longer periods. GK Per, for example, is
an Intermediate Polar with {\it hard} X-ray variations
\citep{watson} that resemble those of V4743 Sgr. It has an orbital
period of nearly 2 days.

Clearly, it is important to determine the binary period, orbital
velocity, and the radius of the cooler component. The long decline
time could indicate the eclipse of extended emitting regions but
this does not seem consistent with the short oscillation period of
22\,min, which implies that the X-ray emitting region is compact.
For the eclipse scenario to be plausible, it must explain why the
harder fraction of the spectrum begins its decline before the
softer fraction of the spectrum.  During an eclipse of an
accretion disk, for example, we expect the outer, cooler, parts to
be occulted first so that the spectrum becomes harder during
ingress, contrary to what is observed.

The rapid oscillations found in the X-ray light curve exhibit the
strongest period at $\sim$22\,min. This period could be the
rotation period of the WD although the complex appearance of the
light curve suggests that there is more than one period present in
the data which argues against rotation.  Drake et al. (2003)
interpreted the oscillations seen in the CHANDRA observation of
V1494 Aql as pulsations because there were multiple periods
present in the light curve of that CN.  Given the uncertainties,
in addition, our analysis implies that the WD is slightly hotter
when it is brighter which implies pulsation and not rotation
because it seems too early in the evolution of the outburst for
accretion at the poles to have started.  Nevertheless, the large
amplitude of the oscillations suggests that it could be rotation
if we can determine a source of asymmetric light from the system
this early in the outburst.  Therefore, determining the cause of
the oscillations will require more data both from satellites
(X-rays) and the ground. Finally, we mention that the BeppoSAX
light curve of V382 Vel, obtained during the SSS phase, was also
found to be highly variable \citep{orio02}, implying that novae
show light variations in the SSS phase. However the cause does
not, necessarily, have to be the same.

The shape of the X-ray spectrum of V4743 Sgr is consistent with
the interpretation that we are viewing the optically thick
atmosphere of a hot WD (see also Paerels et al. 2001).  Both the
continuum shape and the luminosity (determined from the flux and a
distance $\sim$6\,kpc), imply that we have found this nova in the
SSS phase of evolution.  SSS emission is believed to originate
from the hot, optically thick, atmosphere of a WD with ongoing
nuclear burning near the surface \citep{kahab97}.  The compact
source size implied by the 22\,min oscillation is compatible with
this view.  The WD in V4743 Sgr is still hot and luminous because
there is ongoing CNO burning in the accreted layers six months
after discovery.  We note that although both are considered to be
SSS, the spectrum that we show in Figure~\ref{spec} only qualitatively
resembles the X-ray spectrum of CAL 83 obtained by XMM (see
Paerels et al. 2001).  Interesting features of our spectrum are
the widths of the absorption features and the wavelength shifts of
the few identifiable lines which suggest a velocity of
-2400\,km\,s$^{-1}$.  The blue shift implies an ongoing outflow of
material from the WD which may not be consistent with the presence
of the oscillations if they are caused by pulsation.  In addition,
the spectra extracted from the maxima of the oscillations are
slightly harder than those extracted from the minima.  However,
the equivalent widths of the absorption features are little
changed.  While these lines should be formed close to the surface
of the hot WD and change during the oscillations, we cannot tell
where they are formed until we have done a stellar atmosphere
analysis (in progress).  Finally, the spectrum extracted from
after the decline shows only emission lines, suggesting that we
are observing the optically thin ejected gases.

The X-ray light curve and the LETGS spectrum of V4743 Sgr have
produced new puzzles which, when understood, will provide
important insights into the mechanism of nova explosions. The data
presented here are the first part of a CHANDRA observation program
that will observe this nova every two months during summer 2003.
More detailed analysis is necessary and will be carried out when
the observational program is finished. Hopefully, we will see
another eclipse, or even the egress phase in one of the future
observations. Determination of the orbital period and spatial
distribution of the plasma will then be possible.

\acknowledgments

J.-U.N. acknowledges support from DLR under 50OR0105. S.
Starrfield received partial support from NSF and NASA grants to
ASU. JJD and YB were supported by NASA contract NAS8-39073 to the
{\em CHANDRA X-ray Center}. SS, RMW, PHH, JWT, RDG, and CEW were
partially supported by grants from the CHANDRA X-ray Center to
their various institutions. Finally, we would like to thank the
mission planners and staff of the CXC for flawless scheduling and
execution of this ToO program.

\end{document}